**Why Synthetic Control estimators are biased and what to do about it:**

**Introducing Relaxed and Penalized Synthetic Controls**

Oscar Engelbrektson

Minerva University

Advised by Prof. Alexis Diamond

Spring 2021

# Abstract

This paper extends the literature on the theoretical properties of synthetic controls to the case of non-linear generative models, showing that the synthetic control estimator is generally biased in such settings. I derive a lower bound for the bias, showing that the only component of it that is affected by the choice of synthetic control is the weighted sum of pairwise differences between the treated unit and the untreated units in the synthetic control. To address this bias, I propose a novel synthetic control estimator that allows for a constant difference of the synthetic control to the treated unit in the pre-treatment period, and that penalizes the pairwise discrepancies. Allowing for a constant offset makes the model more flexible, thus creating a larger set of potential synthetic controls, and the penalization term allows for the selection of the potential solution that will minimize bias. I study the properties of this estimator and propose a data-driven process for parameterizing the penalization term.[1,2]

(*JEL* C21, C23, C31, C33, C38)

---

[1] *This paper is supplemented by a Python library,* <u>SyntheticControlMethods</u>, *implementing all methods covered in this paper, including the original synthetic control—something which previously did not exist in Python. As of March 20, 2021 the package has been downloaded 12,424 times.*

[2] *I would like to express my gratitude to my advisor Prof. Alexis Diamond, for providing guidance and encouragement throughout the course of this project—and for teaching me about synthetic controls in the first place. I would also like to express a special thank you to Anna Sandros Hansson, Taha Bouhoun, Christopher Wolfram, Johannes Halkenhäußer, Juan Castro Fernandez, Aaron Moralez Shildrick, Erkin Polat, Mau Urdaneta and Eduardo Gomez Videla for helpful discussions and putting up with my ceaseless drivel about synthetic controls.*

# 1. Introduction

In a series of seminal papers Abadie, Diamond, & Hainmueller (2015), Abadie, Diamond, & Hainmueller (2010), and Abadie & Gardeazabal (2003) introduce the synthetic control method, an increasingly popular approach to estimating the effect of a treatment in comparative case studies. Describing the synthetic control method Susan Athey and Guido Imbens (2017) state that "the simplicity of the idea, and the obvious improvement over the standard methods, have made this a widely used method in the short period of time since its inception", making it "arguably the most important innovation in the policy evaluation literature in the last 15 years". The synthetic control method has seen widespread use across a diverse set of fields: from economics to biomedical sciences, to social science, to engineering (Abadie, 2020). In general, synthetic controls are used in settings where a single, or a few, aggregate units have been subjected to treatment and only a limited number of untreated units are available for comparison. For example, Abadie et al. (2015) use it to estimate how West Germany's economy was affected by the German Reunification in 1990, using similarly developed countries as potential untreated units for comparison. To name a use-case from business, synthetic controls are used at Uber to measure the effect on user engagement and user satisfaction of market-wide changes to its product, using unaffected markets as potential comparison units (Jones & Barrows, 2019).

Conceptually, the objective of the synthetic control method is to create a synthetic copy of the treated unit that never received the treatment by combining untreated units. More precisely, the synthetic control method provides a data-driven process for selecting the weighted average of the untreated units that most closely resembles the pre-treatment characteristics of the treated unit. Importantly, these pre-treatment characteristics on which the similarity of units is evaluated includes both the outcome of interest and the observed predictors of the outcome of interest. Consider for example Abadie et al. (2010), which studies the effect of Proposal 99, a large-scale tobacco control program that California implemented in 1988. Their outcome of interest is per capita cigarette consumption and they include as explanatory variables, inter alia, the average retail price of a pack of cigarettes, per capita income and per-capita beer consumption.

In this paper, I show that synthetic control estimators are generally biased if the relationship between outcome variable and the explanatory variables studied is non-linear. This is a meaningful contribution as it is important for researchers to be aware that if the relationship is not linear, biased estimates will result—especially because the true relationship is often unknown in empirical applications. Furthermore, I show that this bias increases with, amongst other things, the weighted sum of the absolute unitwise discrepancies between the treated unit and the untreated units in the synthetic control.

To explain what this means, it might be helpful to conceptualize the unit-weighting performed by the synthetic control method as "mixing" untreated units into a synthetic control unit. Under this conceptualization, the objective of the synthetic control method can be thought of as finding weights such that *after* mixing them into a synthetic control, the untreated units are maximally similar to the treated unit. What my analysis then shows is that if the relationship between the outcome variable and the explanatory variables is non-linear, then (a) the synthetic control is generally biased even if the untreated units match the pre-treatment characteristics of the treated unit *after* mixing them into a synthetic control unit and (b) the magnitude of this bias becomes increases with the difference between the treated unit and the individual untreated units inside the synthetic control *before* you mix them.

Following this realization, I propose a novel synthetic control estimator which strives to maximize the similarity between the treated unit and the untreated units not only *after* mixing them into a synthetic control (as in ordinary synthetic controls), but also that the untreated units in the synthetic control be maximally similar *before* you mix them. Of course, the importance of similarity before mixing depends on the degree of non-linearity in the specific context that the synthetic control is being applied. If the relationship is highly non-linear, a small difference before mixing is imperative; whereas if the relationship is perfectly linear, the difference before mixing does not matter. To address this issue, I propose a data-driven process for determining the relative importance of similarity before versus after mixing for each specific application.

The paper is structured as follows. Section 2 formalizes the theoretical setting and notation used throughout the paper and provides a formal definition of synthetic controls. Section 3 summarizes previous work on the theoretical properties of the synthetic control estimator and then extends this literature to the general case of non-linear generative processes, proving that synthetic control estimators are generally biased in such settings. Section 4 proposes a novel synthetic control estimator designed to minimize this bias—the *relaxed and penalized synthetic control*—and analyses its properties. Section 5 evaluates the performance of this novel estimator using empirical data. The final section concludes and describes open areas for future research.

# 2. Formal introduction to Synthetic Controls

## 2.1. Notation and the potential outcomes framework

Consider J+1 units observed in time periods t = {1,2,...,T}. Without loss of generality, assume the unit at index 1 is the only treated unit, the remaining J units at indices {2,3,...J+1} are untreated. In keeping with Abadie et al. (2010) we define $(Y_{it}^N, Y_{it}^I)$ to be the potential outcome for unit $i$ at time $t$ in the absence of treatment and in the presence of treatment, respectively. Additionally, for each unit $i$ at time $t$, along with the outcome $Y_{it}$, we also observe a set of $k$ predictors of the outcome. We define $T_0$ to be the time period of the intervention. Furthermore, we define the treatment effect to be the difference in potential outcomes for unit $i$ at time $t$. We take $D_{it}$ to be the indicator variable representing the treated status of unit $i$ at time $t$, such that $D_{it}$ = 1 if $i$=1 and $t > T_0$, and $D_{it}$ = 0 otherwise. Consequently, the observed outcome $Y_{it}$, for any unit at any time can be expressed as,

$$Y_{it} = Y_{it}^N + \tau_t D_{it}$$

where $\tau_t$ is the treatment effect. The objective is to estimate the treatment effect on the treated,

$$\tau_{1t} = (Y_{1t}^I - Y_{1t}^N) \text{ for } t > T_0 \qquad (1)$$

i.e. the difference in potential outcomes for the treated unit in the post-treatment period. However, as $Y_{1t}^N$ is never observed in the post-treatment period, it must be estimated using available data. Consequently, the objective of the models evaluated in this paper, is to estimate this unobserved quantity–what the outcome of the treated unit *would have been* if it had not received the treatment.

## 2.2 Formal definition of Synthetic Controls

Formally, any weighted average of the untreated units is a synthetic control and can be represented by a convex J × 1 vector of weights $\mathbf{W} = (w_2,...,w_{J+1})$, such that $w_j \in (0,1)$ and $w_2 + \ldots + w_{J+1} = 1$. The rationale for the convexity constraint is that it guards against extrapolation, such that synthetic controls are always weighted averages of the untreated units (Abadie et al., 2010). The objective is to find the $\mathbf{W}$ for which the characteristics of the treated unit are most closely approximated by those of the synthetic control. Let $\mathbf{X_1}$ be a k × 1 vector consisting of the pre-intervention characteristics of the treated unit which we seek to match in the synthetic control. Operationally, each value in $\mathbf{X_1}$ is a pre-treatment average of each covariate for the treated unit, thus k is equal to the number of covariates in the dataset. Similarly, let $\mathbf{X_0}$ be a k × J matrix containing the pre-treatment characteristics for each of the J untreated units. The difference between the pre-treatment characteristics of the treated unit and a synthetic control can thus be expressed as $\mathbf{X_1} - \mathbf{X_0}\mathbf{W}$. We select $\mathbf{W^*}$ to minimize this difference.

$$\min_{W \in R^{n_0}} V| \ |X_1 \ - \ \sum_{i=2}^{J+1} W_i X_i \ | \ |^2 \qquad (2)$$

Where $\mathbf{V}$ is a k × k diagonal, semidefinite matrix where each element $v_{j,j}$ represents the relative importance of the synthetic control reconstructing the treated units value for covariate $j$. Given $W,$ the synthetic control estimate of the outcome of the treated unit in the absence of treatment is

$$Y^N_{1t} = \sum_{i=2}^{J+1} w_i Y_{it} \qquad (3)$$

The treatment effect can thus be computed using equation (1) (Abadie, 2020).

# 3. Theoretical analysis of bias

## 3.1 Previous work

Theoretical analysis side-steps the algorithmic and implementation considerations involved in the selection of **W**, and instead focuses on the bias properties that follow from a given **W**. Abadie et al., (2010) analyze the bias properties of synthetic controls under the assumption that the outcomes are generated by a linear factor model, such that

$$Y^N_{it} = \delta_t + \theta_t Z_i + \lambda_t \mu_i + \varepsilon_{it} \qquad (4)$$

Where $\delta_t$ is a common time trend, $Z_i$ and $\mu_i$ are unit-specific observed and unobserved predictors of the outcome $Y^N_{it}$, respectively. $\theta_t$ and $\lambda_t$ are common, time-varying coefficients on the predictors. That is, the functional relationship between the predictors and the outcome is assumed to be the same for all units, treated and otherwise. Lastly, $\varepsilon_{it}$ is a unit-specific, time-varying noise term with mean zero. Specifically, they study bias in the case where the synthetic control unit perfectly matches the observable pretreatment characteristics of the treated unit. They let $\mathbf{X_1}$ be a vector that consists of the covariates from $\mathbf{Z_1}$ and the pre-treatment outcomes of the treated unit. Similarly, they let $\mathbf{X_0}$ be the matrix containing the same variables for all the untreated units. They show that if the synthetic control perfectly recreates the pretreatment characteristics of the treated unit, such $\mathbf{X_1} = \mathbf{X_0W}$, then the bias of the estimated treatment effect, $\tau_{1t}$ is bounded by the ratio of the magnitude of noise term $\varepsilon_{it}$ to the number of pre-treatment periods. This result can be understood intuitively: given the model from equation (4), a synthetic control that matches observed $\mathbf{Z_1}$ and unobserved $\mu_1$ yields an unbiased estimate of the treatment effect, as the errors are mean zero. By definition, $\mathbf{X_1} = \mathbf{X_0W}$ implies that that

$$Y_{1t}^N = \sum_{i=2}^{J+1} w_i Y_{it} \text{ and } Z_1 = \sum_{i=2}^{J+1} w_i Z_{it} \text{ for all } t \leq T_0.$$ The fit on $\mu_1$ cannot be empirically

evaluated as it is unobserved. However, assume that the synthetic control failed to replicate $\mu_1$, then for $\mathbf{X_1} = \mathbf{X_0W}$ to hold the differences in the noise terms of the treated unit and the synthetic control must exactly offset the effect of imbalance on unobservables. This becomes increasingly unlikely as the number of pre-treatment periods increases, the magnitude of the noise term decreases, and the number of untreated units decreases (Abadie et al., 2010).

Making the same assumptions on the generative process, Ferman and Pinto (2020) extend the analysis to the more realistic case where the pre-treatment fit is only approximate, i.e. $\mathbf{X_1} \approx \mathbf{X_0W}$. They show that for $\mathbf{X_1} \approx \mathbf{X_0W}$ to be possible in practice, typically necessitates that the magnitude of the noise be small.

## 3.2 Clarification of assumptions

It is prudent to highlight some assumptions implicit in the above analyses. Firstly, all the units included in the control group, or at least those which are assigned a non-zero weight in the synthetic control, must share the same functional relationship between outcome and predictors as the treated unit. Secondly, the stable unit treatment value assumption, SUTVA, is implicit.

Specifically, the synthetic control can only provide an unbiased estimate of the outcome in the absence of treatment if none of the untreated units that constitute it were affected by the treatment. Thirdly, the synthetic control can only provide unbiased estimates of the counterfactual outcome in the posttreatment period on the condition that none of the untreated units that constitute it were subjected to some treatment in any of the post-treatment periods that affects the value of the some predictors.

### 3.3 Non-linear outcome models

Previous literature that engages in analysis of the bias properties of the Synthetic Control do so assuming the outcome is a linear function of the predictors (Abadie et al., 2010; Abadie, 2020; Ben-Michael et al., 2020; Ferman & Pinto, 2018). The following section shows that the Synthetic Control is generally biased if the outcome model is non-linear. As the true functional relationship between the outcome and predictors is unknown in many applications–and thus plausibly non-linear–this is an insight with important practical implications for analysis.

To illustrate this argument, consider the most basic scenario with no unobserved predictors, no noise and no time-trend. In this scenario, the observed pre-treatment predictors of a unit can explain a hundred percent of the variance in that unit's pretreatment outcomes. In the setting where the generative model is linear, selecting and evaluating $\mathbf{W}$ should be trivial. If $\mathbf{X_1} = \mathbf{X_0}\mathbf{W}$, then the estimated treatment effect is guaranteed to be unbiased with zero variance. Furthermore, so long as the outcome of the treated unit is not more extreme than all the untreated units, the existence of such a solution is guaranteed. However, imagine instead that the outcome was a quadratic function of the observed covariates,

$$Y_{it}^N = \theta_t Z_i^2 \qquad (5)$$

In this scenario, the observed pre-treatment predictors of a unit still explain all of the variance in the unit's pretreatment outcomes. However, in general, there exists no $\mathbf{W}$ such that $\mathbf{X_1} = \mathbf{X_0 W}$. To see why, recall that $\mathbf{X_1} = \mathbf{X_0 W}$ implies that $Y_{1t}^N = \sum_{i=2}^{J+1} w_i Y_{it}$ and $Z_1 = \sum_{i=2}^{J+1} w_i Z_i$ for the pre-treatment period. Substituting $Y_{it}^N = \theta_t Z_i^2$ from equation (5) and dividing by $\theta_t$, we get that there must be a solution $\mathbf{W}$ such that the following equalities hold simultaneously,

$$Z_1 = \sum_{i=2}^{J+1} w_i Z_i \qquad \text{and} \qquad Z_1^2 = \sum_{i=2}^{J+1} w_i Z_i^2$$

However, this is generally impossible given the convexity constraints on $\mathbf{W}$.[3] Note that the quadratic exponent can be exchanged for any exponent except 1 without affecting the veracity of the above statement–or any other non-linear function, for that matter (see appendix A). The exception to this is if there is a single untreated unit that has the exact same pretreatment characteristics as the treated unit. However, in that scenario, there is no need for the Synthetic Control in the first place–a normal, non-synthetic control already exists!

**Example 1:** Consider a simple numerical example where the outcome, $Y_i$ is the square of a single observed covariate, $Z_i$. Suppose, there is one treated unit with $Z_1 = 2$, $Y_1=4$ and two untreated units with $Z_2 = 1$, $Y_2=1$ and $Z_3 = 3$, $Y_3=9$, respectively. This simple setting depicted in Figure 1.

---

[3]The convexity constraint requires that all unit-weights must be non-negative and together sum to one. That is, $\mathbf{W} = (w_2,...,w_{J+1})$, such that $w_j \in (0,1)$ and $w_2 + ... + w_{J+1} = 1$.

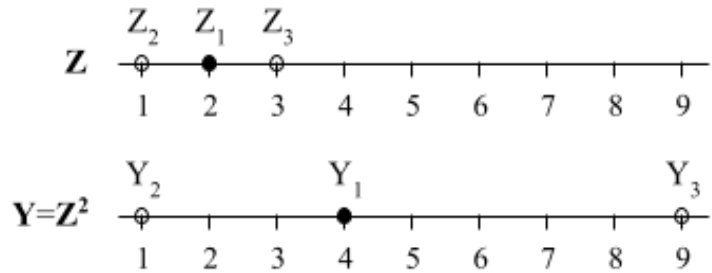

*Figure 1*: A simple example

Notice that the covariate value of the treated unit is exactly in the middle of $Z_2$ and $Z_3$, such that $Z_1$ can be perfectly reconstructed by assigning equal weight to both untreated units, i.e.

$Z_1 = w_2(1) + w_3(3)$ with $\mathbf{W} = \{0.5, 0.5\}$. However, notice that this configuration of weights fails to reconstruct the outcome of the treated unit, as $0.5(1) + 0.5(9) = 5 \neq 4$. Similarly, the weights that reconstruct the outcome of the treated unit, $\mathbf{W} = \{0.625, 0.375\}$, fail to reconstruct the covariate $Z_1$ as $0.625(1) + 0.375(3) = 1.75 \neq 2$. Any set of intermediary weights, will trade off error on $Z_1$ for error on $Y_1$, without perfectly matching. In summary, if the outcome is a nonlinear function of the covariates, it is possible to reconstruct covariates or the outcome, but impossible to do both simultaneously.

At this point, we consider the implications of adding back in the noise term to equation (5). This is important because, in the real world, if an outcome can be perfectly predicted from observed variables it is unlikely to warrant the use of the Synthetic Control method. Consider,

$$Y_{it}^N = \theta_t Z_i^2 + \varepsilon_{it} \quad (6)$$

This does nothing to alleviate the fundamental problem posed by non-linearity. However the introduction of noise creates the possibility for spurious solutions to $\mathbf{X_1} = \mathbf{X_0}\mathbf{W}$.

**Example 2**: Consider an extension of the setting in example 1. Again suppose there are three units with the same covariate values as in example 1: $Z_1 = 2$, $Z_2 = 1$, and $Z_3 = 3$. However, this time we introduce an error term, such that the outcomes $Y_i$ are given by equation (6). To illustrate how this addition introduces the possibility of spurious solutions, assume the values of the three error terms are $\varepsilon_1 = 1$, $\varepsilon_2 = 1$, $\varepsilon_3 = $ -1. Consequently, the outcomes are $Y_1 = 5$, $Y_2 = 2$, $Y_3 = 8$. This setting is depicted in figure 2.

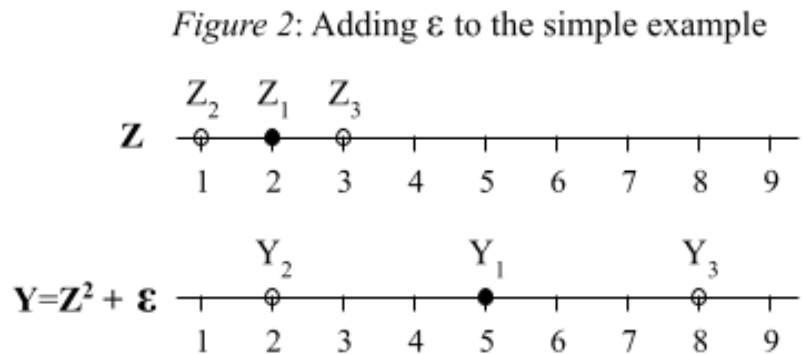

*Figure 2*: Adding ε to the simple example

Recall that $\mathbf{W} = \{0.5, 0.5\}$, resulted in a match on $Z_1$, but a mismatch of 1 on the outcome $Y_1$. As $Z_1$ is unchanged, $\mathbf{W} = \{0.5, 0.5\}$ still results in a match thereon. In this case, it also results in a perfect match on the outcome as $0.5(2) + 0.5(8) = 5$. It looks like a perfect synthetic control, but we know it is not–it was achieved by exploiting the particular realizations of the error term, not by successfully reconstructing the treated unit. Indeed, we know from example 1 that a perfect synthetic control is impossible in this non-linear setting without the error term.

In practice then, if we observe $\mathbf{X_1} = \mathbf{X_0W}$ in the non-linear case, it was probably achieved by overfitting on the noise terms. For completeness, it should be noted that making some observed predictors considered in equation (6) unobserved instead, and thereby reintroducing $\lambda_t \mu_i$, does nothing to alleviate the fundamental problem posed by non-linearity either. Similarly, the problem cannot be avoided fitting only to outcomes by excluding covariate values from $\mathbf{X_1}$ and $\mathbf{X_0}$. This would be practically equivalent to all covariates being unobserved, yielding the generative model

$$Y_{it}^N = \lambda_t \mu_i^2 + \varepsilon_{it} \qquad (7)$$

and the logic for bias in equation (6) is thus directly applicable as the conceptual difference between $\mu_i$ and $Z_i$ is merely whether the covariates they contain are in $\mathbf{X}$.

### 3.3 What the bias scales with - deriving a lower bound for the bias

Consider a generative model like the one in equation (4), but which does not make assumptions on the functional form of the relationship between the outcome and covariates,

$$Y_{it}^N = \delta_t + \Phi_t(Z_i) + \phi_t(\mu_i) + \varepsilon_{it} \qquad (8)$$

Where $\Phi_t(\bullet)$ and $\phi_t(\bullet)$ are some shared, time-varying functions which map the observable and unobservable covariates to the outcome, respectively. The synthetic control can then be expressed as,

$$\sum_{i=2}^{J+1} w_i Y_{it} = \delta_t + \sum_{i=2}^{J+1} w_i \Phi_t(Z_i) + \sum_{i=2}^{J+1} \phi_t(\mu_i) + \sum_{i=2}^{J+1} \varepsilon_{it}$$

Consequently, the difference between the treated unit and the synthetic control is,

$$Y_{1t}^N - \sum_{i=2}^{J+1} w_i Y_{it} = \Phi_t(Z_1) - \sum_{i=2}^{J+1} w_i \Phi_t(Z_i) + \phi_t(\mu_1) - \sum_{i=2}^{J+1} w_i \phi_t(\mu_i) + \varepsilon_{1t} - \sum_{i=2}^{J+1} w_i \varepsilon_{it}$$

If we find a synthetic control with good pretreatment fit such that $\mathbf{X_1 = X_0 W}$, it still implies that that $Y_{1t}^{N} = \sum_{i=2}^{J+1} w_i Y_{it}$ and $Z_1 = \sum_{i=2}^{J+1} w_i Z_i$ for all $t \leq T_0$. Therefore,

$$(\Phi_t(Z_1) - \sum_{i=2}^{J+1} w_i \Phi(Z_i)) + (\phi_t(\mu_1) - \sum_{i=2}^{J+1} w_i \phi_t(\mu_i)) + (\varepsilon_{1t} - \sum_{i=2}^{J+1} w_i \varepsilon_{it}) = 0$$

And rearranging gives,

$$\Phi_t(Z_1) - \sum_{i=2}^{J+1} w_i \Phi(Z_i) = \phi_t(\mu_1) - \sum_{i=2}^{J+1} w_i \phi_t(\mu_i) + \varepsilon_{1t} - \sum_{i=2}^{J+1} w_i \varepsilon_{it} \qquad (9)$$

Expressed verbally, any difference between $\Phi_t(Z_1) - \sum_{i=2}^{J+1} w_i \Phi_t(Z_i)$ must be exactly offset by

$\phi_t(\mu_1) - \sum_{i=2}^{J+1} w_i \phi_t(\mu_i) + \varepsilon_{it} - \sum_{i=2}^{J+1} w_i \varepsilon_{it}$. In the best case, $\phi_t(\mu_1) - \sum_{i=2}^{J+1} w_i \phi_t(\mu_i) = 0$,

meaning that unobserved covariates are perfectly matched and contribute nothing towards bias. In that case, the entire difference in the left hand side of equation (9) was offset by overfitting to noise.

Consequently, if $\mathbf{X_1} = \mathbf{X_0}\mathbf{W}$, a lower bound on the bias of the synthetic control is given by

$$\Phi_t(Z_1) - \sum_{i=2}^{J+1} w_i \Phi_t(Z_i) \qquad (10)$$

Equation (10) increases with three things. The degree of non-linearity of $\Phi_t(\bullet)$, the magnitude of $Z_1$, and the weighted sum of pairwise differences between the observable covariates of the treated unit and the untreated units, $\sum_{i=2}^{J+1} w_i ||Z_1 - Z_i||^2$. I will provide an example to illustrate each of the three drivers of bias.

**Example 3: Degree of non-linearity**

Let us once again return to the setting of example 1. Recall that the covariates were $Z_1=2$, $Z_2=1$, $Z_3=3$ and the outcome was $Y_i = Z_i^2$. That is, $\Phi(Z_i) = Z_i^2$ (for notational simplicity we drop the time subscript for the remainder of this section). In this setting, $\mathbf{W}=\{0.5, 0.5\}$ resulted in a perfect match on the observable covariates, $Z_1 = \sum_{i=2}^{J+1} w_i Z_{it}$, but a mismatch of 1 on the outcome, $|\Phi(Z_1) - \sum_{i=2}^{J+1} w_i \Phi(Z_i)| = 1$. Consider now what were to happen if we changed the strength of the non-linearity such that $\Phi(Z_i)=Z_i^3$, keeping everything else constant. This setting is depicted

in Figure 3:

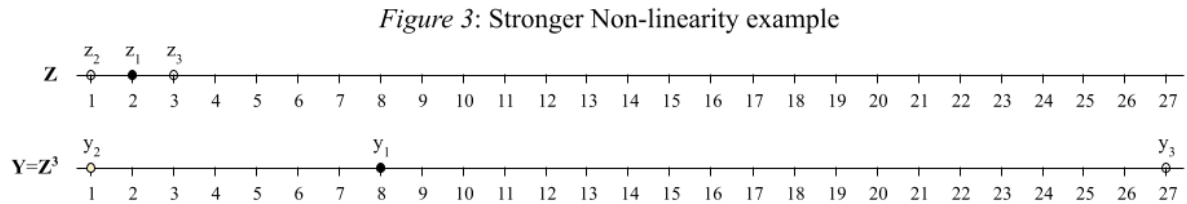

*Figure 3*: Stronger Non-linearity example

Here, **W**={0.5, 0.5} still results in a perfect match on the observable covariates, as they are

unaffected. However, $|\, \Phi(Z_1) - \sum_{i=2}^{J+1} w_i \Phi(Z_i)\,|$ has increased from 1 to 6 as

8 - 0.5(1) + 0.5(27) = 6. This effect would have been even more extreme if the non-linearity of

$\Phi(\bullet)$ was even stronger. Conversely, if $\Phi(\bullet)$ was closer to being linear, the bias would decrease.

In fact, in the edge case where $\Phi(\bullet)$ is linear, $\sum_{i=2}^{J+1} w_i \Phi(Z_i) = \Phi(\sum_{i=2}^{J+1} w_i Z_i)$, and it is sufficient that $Z_1 = \sum_{i=2}^{J+1} w_i Z_{it}$ for observable covariates to contribute nothing towards the bias.

As Abadie et al. (2010) study the case where $\Phi(\bullet)$ is linear and $\mathbf{X_1 = X_0 W}$, which implies that

$Z_1 = \sum_{i=2}^{J+1} w_i Z_{it}$, this explains why the bias bound they derived does not depend on observable

covariates. Thus, in the edge case where $\Phi(\bullet)$ is linear, the results implied by the framework in this section are entirely consistent with Abadie et al. (2010). This is worth noting because it illustrates the results of this analysis are not contradictory to theirs, rather the differences stem from the weaker assumptions I make about the relationship between outcome and covariates.

**Example 4: Magnitude of observable covariates**

Imagine the same setting as in example 3, but where we have shifted each covariate $Z_i$ 10 units along the number line. In this scenario $Z_1$=12, $Z_2$=11, and $Z_3$=13. Here, $\Phi(Z_i) = Z_i^3$ and the pairwise difference between each of the units is unchanged, i.e. $\|Z_1 - Z_2\|^2 = 1$ and $\|Z_1 - Z_3\|^2 = 1$. Similarly, **W**={0.5, 0.5} still results in a perfect match on the observable covariates, as 12 - 0.5(11) + 0.5(13) = 0. However, $Y_1$=1728, $Y_2$=1331, and $Y_3$=2197. As a result, $|\Phi(Z_1) - \sum\limits_{i=2}^{J+1} w_i \Phi(Z_i)|$ has changed from 1 to 36, as 1728 - 0.5(1331) + 0.5(2197) = 36. In general, so long as $\Phi(\bullet)$ is nonlinear, equation (10) increases with the magnitude of the $Z_1$, all else equal.

**Example 5: Weighted sum of pairwise differences**

Imagine the same setting as in example 1, but this time we have introduced a new unit $Z_4$=4, $Y_4$=16. This setting is depicted in Figure 4.

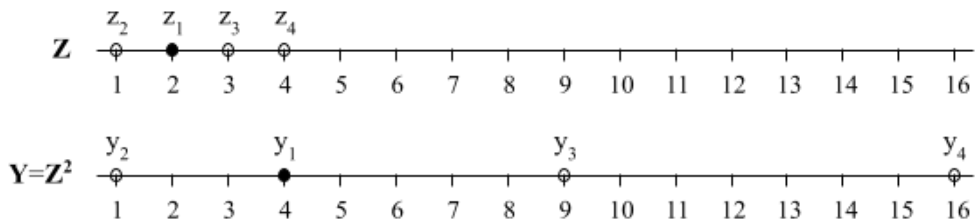

*Figure 4*: Pairwise difference example

The original solution assigning 50% weight to units 2 and 3 respectively exists in

$\mathbf{W_1}=\{\frac{1}{2}, \frac{1}{2}, 0\}$, and still results in a perfect match on covariates but mismatch of 1 on the

outcome. In this solution, the weighted sum of pairwise discrepancies is $\sum_{i=2}^{J+1} w_i||Z_1 - Z_i||^2 =$

$0.5(2-1)^2+0.5(2-3)^2 + 0(2-4)^2 = 1$. However, with the addition of another untreated unit, it is also

possible to match on covariates with $\mathbf{W_2}=\{\frac{2}{3}, 0, \frac{1}{3}\}$, as $2 - \frac{2}{3}(1) + 0(3) + \frac{1}{3}(4) = 0$. Note that with $\mathbf{W_2}$ $\sum_{i=2}^{J+1} w_i||Z_1 - Z_i||^2 = \frac{2}{3}(2-1)^2+ 0(2-3)^2 + \frac{1}{3}(2-4)^2 = 2$, which is an increase of 1. For this reason, this new solution increases the mismatch on outcome $|\Phi(Z_1) - \sum_{i=2}^{J+1} w_i\Phi(Z_i)|$ from 1 to 2, as $4 - \frac{2}{3}(1) + 0(9) + \frac{1}{3}(16) = 2$. More generally, there is an infinite number of solutions

between the original solution $\mathbf{W_1}=\{0.5, 0.5, 0\}$ and $\mathbf{W_2}=\{\frac{2}{3}, 0, \frac{1}{3}\}$. However, the more

weight we assign to $Z_4$, the more the pairwise error increases, and with it best-case bias

$|\Phi(Z_1) - \sum_{i=2}^{J+1} w_i\Phi(Z_i)|$.

To conclude, notice that both the magnitude of the observable covariates in $Z_1$ and the functional form of $\Phi(\bullet)$ are immutable facts of the data, and are therefore beyond the control of the researcher. Once a dataset has been collected, there is nothing a researcher can do to affect either. Consequently, by process of elimination, the only means by which a researcher can decrease the lower bound of the bias of their synthetic control application is by minimizing the weighted sum of pairwise differences between the treated unit and untreated units.

**3.4 Implications for the synthetic control method - modifying the covariate balance table**

Both the first and the third cause of bias in the non-linear case are directly determined by the data, and are as such beyond the control of the researcher. The second, however, is affected by the choice of $\mathbf{W}$, which we do control. The bias increases as some function of the pairwise distances between the treated unit and the untreated units in the synthetic control.

When using the synthetic control method, it is common to evaluate the quality of the pretreatment fit of the synthetic control by comparing the weighted average of observed pretreatment predictors in the synthetic control to those of the treated unit. That is, one column which displays $\mathbf{X_1}$ and a second which displays $\mathbf{X_0W}$. However, the synthetic control unit is not a unit in itself, merely a weighted average of units. As such, we risk falling prey to a kind of anti-reductionist fallacy, whereby we believe the synthetic control unit is more than simply the sum of its parts. It's not, by definition. In other words, achieving good pre-treatment fit on the pre-treatment predictors is necessary but not sufficient for determining the quality of the synthetic control. To address this, I propose to include an additional column in covariate balance tables which shows the weighted average of absolute pairwise discrepancies between the units in the synthetic control and treated unit for each covariate. Interpreting such a table, one would want to observe similar values in each row of the two first columns, and small numbers, relative to the magnitude of the covariate values, in the third column. Large numbers in the third column would imply increased risk for biased estimates.

|  | California | Synthetic California | WMAPE | Importance |
|---|---|---|---|---|
| **cigsale** | 117.66 | 117.67 | 56.58 | 0.16 |
| **lnincome** | 10.02 | 9.71 | 0.32 | 0.21 |
| **beer** | 24.45 | 22.55 | 8.71 | 0.31 |
| **age15to24** | 0.18 | 0.19 | 0.01 | 0.20 |
| **retprice** | 63.82 | 61.11 | 2.71 | 0.12 |

*Table 1.* Average values of outcome and observed covariates of California (first column) and the synthetic control unit for California (second column). The synthetic control unit closely matches the average pre-treatment values of California for each of the five variables, as evidenced by the similarity of the values in the first and second column of each row. The third column, WMAPE, is showing the weighted sum of absolute pairwise differences between the California and the untreated units that constitute Synthetic California. Smaller values indicate that the control units were similar to the treated unit before mixing. If the value is 0, it means that each of the control units assigned non-zero weight in the synthetic control have exactly the same value for that covariate as the treated unit. In general, whether or not the values of the third column are large should be evaluated with reference to the magnitude of the values in the first column. In this example, we can say that the average unit in the synthetic control is on average within ±50% of the outcome ("cigsale") of California, as $0.5(117) \approx 56$. The last column, "Importance", shows the leading diagonal of **V,** indicating the relative importance of achieving balance on the corresponding variable determined by the synthetic control solution. *This table was generated using the SyntheticControlMethods package (See appendix B).*

# 4. Penalized and Relaxed Synthetic Controls

## 4.1 Introducing Penalized and Relaxed Synthetic Controls

This section of the paper introduces a penalized synthetic control estimator designed to alleviate challenges described in the previous section. Given a positive penalization constant $\lambda$ and a covariate importance matrix $\mathbf{V}$, the penalized synthetic control weights $\mathbf{W(V, \lambda)^*}$ solve

$$\min_{W \in R^{n_0}} V|\ |X_1^{\Delta}\ -\ \sum_{i=2}^{J+1} W_i X_i^{\Delta}\ |\ |^2\ +\ \lambda V \sum_{i=2}^{J+1} W_i|\ |X_1\ -\ X_i\ |\ |^2$$

(11)

Where the $\Delta$ superscript indicates the first difference has been taken. That is, whilst $X_1$ represents the average value of each predictor of the treated unit over the pre-treatment period, $X_1^{\Delta}$ represents the average *change* each predictor of the treated unit over the pre-treatment period. The penalized differenced synthetic control estimates are then

$$\tau_{1t} = Y_{1t}^I\ -\ \sum_{i=2}^{J+1} w_i\ Y_{it}\ \ \text{for } t > T_0 \tag{12}$$

Equation (11) consists of two terms, the first relaxes the original synthetic control estimator from equation (2), whereas the second term penalizes it, hence the name. It is the first term of equation (11) that enables the addition of a constant offset $\alpha$ in equation (12). I go into detail about the relaxation term in section 4.2.

The second term penalizes the pairwise discrepancies between the pre-treatment characteristics of the treated unit $X_1$ and each individual untreated unit. This penalization term is identical to the one introduced in Abadie and L'Hour (2020). The rationale for its inclusion and the context for which the estimator is developed, however, is quite different. Abadie and L'Hour (2020) introduce the penalization term as a means to ensure sparsity and uniqueness of solutions to the synthetic control in settings with many treated units. Specifically, they show that if $\mathbf{X_1}$ lies inside the convex hull of $\mathbf{X_0}$, then there is an infinite set of synthetic control weights $\mathbf{W}$ which perfectly reconstruct the pre-treatment characteristics of the treated unit. In this scenario, the penalization term serves as a tie-breaker to ensure a single solution is preferable to all others. Conversely, if $\mathbf{X_0}$ is outside the convex hull of $\mathbf{X_1}$, then the optimal solution to the synthetic control is unique and given by the projection of $\mathbf{X_0}$ onto the convex hull of $\mathbf{X_1}$. They go on to note that

*"One practical consequence of the curse of dimensionality is that, even for a moderate number of matching variables, each particular treated unit is unlikely to fall in the convex hull of the untreated units, especially if the number of untreated units is not very large. As a result, lack of uniqueness is rarely an issue in settings with one or a small number of treated units and, if it arises, it can typically be solved by ad-hoc methods, like increasing the number of covariates or by restricting the donor pool to units that are similar to the treated units. In settings with many treated and many untreated units, non-uniqueness may be an important consideration and a problem which is harder to solve." (Abadie and L'Hour, 2020)*

Thus, viewed from the lens of non-uniqueness, there is no compelling reason to include a penalization term in the setting with a single treated unit. As shown in the previous section of this paper, however, from the perspective of bias there is a clear and compelling reason to penalize the pairwise discrepancy. Nevertheless, including a penalization term in the objective function is only useful to the extent that it shifts the resultant **W** towards a less biased synthetic control, which is only possible if there are multiple potential solutions which achieve good pre-treatment fit. This is where relaxation of the Synthetic Control comes into play.

## 4.2 The case for relaxation

As hinted in the previous subsection, the rationale for creating a more flexible estimator is to ensure there are multiple synthetic control weights which reconstruct the pre-treatment characteristics of the treated unit. This would allow the penalization term to select from the set of possible synthetic control weights, the one that minimizes bias. Phrased geometrically, the aim is to define an analogous problem where the convex hull of the untreated units is larger, such that it is probable for a single treated unit to be contained therein. The remainder of this subsection explains how the first term of equation (11) operationalizes this aspiration.

Firstly, note that the first term of equation (11) differs from the original synthetic control formulation in equation (2) only by substituting $X_1$ for $X_1^{\Delta}$ and $X_0$ for $X_0^{\Delta}$. Whilst $X_1$ represents the average value of each predictor of the treated unit over the pre-treatment period, $X_1^{\Delta}$ represents the average change. Consequently, **V** and $\lambda$ are chosen such that **W(V, $\lambda$)\*** from equation (11) results in a synthetic control that most closely reconstructs the changes in the outcome of the treated unit over the pre-treatment period, $Y_{1t}^{\Delta}$—as opposed to the outcome itself $Y_{1t}$. Once we have chosen such a **W(V, λ)\***, we apply the weights to the original, untransformed untreated group outcomes from the pre-treatment period. Conditional on achieving a good fit in the previous steps, this should yield a synthetic control that changes the same way as the treated unit over the pre-treatment period, but differs from it by some constant offset, $\alpha$. That is, the outcomes have a parallel trend, to borrow from Difference-in-Differences terminology. $\alpha$ can then be computed as the average pre-treatment difference between synthetic control and treated unit in the pre-treatment period, allowing for treatment effect to be estimated by equation (12).

The key realization that underlies this modification to the original synthetic control is that the convex hull of the changes of the control group nests the convex hull of the real values of the control group. To see this, notice that by choosing weights to minimize equation (2) we are fitting to both the outcome value and—by fitting to the value at every time—the change of the outcome of the treated unit. Consequently, the traditional synthetic control method necessitates that the treated unit be inside the convex hull of the outcome variable of the untreated units, for every point in the pre-treatment period, and therefore also that it be inside the convex hull of the changes of the outcome variable. As such, the Synthetic Control is fitting the trajectory of the treated unit only implicitly, by fitting to the value at every time period. This approach does so explicitly, whilst disregarding the real values of the treated unit outcome in solving for **W**, thereby achieving the desired outcome of increasing probability that the treated unit is contained within the relevant convex hull of the untreated units.

Lastly, it is worth noting that other choices of relaxation are possible, such as directly adding a constant to the optimization or relaxing the convexity constraint on **W**. However, I favoured this approach because it increases the flexibility of the estimator whilst preserving its geometric interpretation of the synthetic control solutions.

## 4.3 Estimation

Given a choice of **V** and λ, solving for **W(V, λ)\*** from equation (11) is a constrained quadratic optimization problem, with the constraints being the convexity of **W**. The technique for selecting **V** and λ proposed in this section can be thought of as an extension of the technique for selecting **V** proposed in Abadie et al. (2015). The entire estimation procedure is operationalized as follows:

1. Divide the pre-treatment period into a training period and a validation period with S and L periods, respectively. Let the validation period be the L periods directly preceding the treatment period, and the training period be the S periods directly preceding the validation period.

2. Compute $X_1^{\Delta}$ and $X_0^{\Delta}$ using the data from the training period.

3. Choose **V** and λ to minimize the mean square prediction error during the validation period:

$$\sum_{t=S}^{S+L} (Y_{1t}^{\Delta} - \sum_{i=2}^{J+1} w(V, \lambda)_i^* \, Y_{it}^{\Delta})^2$$

4. Given **V** and $\lambda$ from **3.** compute **W(V, $\lambda$)\*** with $X_1^{\Delta}$, $X_0^{\Delta}$, $X_1$, and $X_0$ from the validation period.

5. Using **W(V, $\lambda$)\*** from **4.**, solve for $\alpha$ using the validation period outcomes, where $\alpha$ is the average difference between the outcomes of the synthetic control and treated unit:

$$\alpha = \frac{1}{S}\sum_{t=1}^{S}(Y_{1t} - \sum_{i=2}^{J+1} w(V, \lambda)_i^* Y_{it}) \qquad (13)$$

The treatment effect can then be estimated by equation (12).

### 4.4 Inference

The placebo-based validity tests—in-time placebos, in-space placebos and variations thereof e.g. pre/post RMSPE ratio histograms—used in the synthetic control method can be directly ported into the differenced synthetic control framework. However, the standard procedure for evaluating covariate balance in synthetic controls applications cannot be applied as the absolute values of the outcome differ by $\alpha$. Therefore, if we compare the absolute values of the covariates we will observe misleading results, as the synthetic control and the treated units have different absolute values. To exemplify the logic, assume some outcome $Y_t^D$, where D represents treatment assignment, is an increasing linear function of the observable covariates and that there are no unobserved characteristics which impact the outcome. Assume also we have $Y_{1t}^I - Y_{1t}^N + \alpha = 0$ for $\alpha > 0$ and $t < T_0$. That is, the pre-treatment outcomes of the treated unit are larger by a constant amount $\alpha$ than synthetic control outcomes. As the difference is exactly constant for every pre-treatment period, we have a perfect pre-treatment fit. Furthermore, because this fit was produced when controlling for all relevant pre-treatment characteristics we know the covariate balance, in actuality, is perfect. However, if we were to make a table showing the absolute values of the covariates of the treated unit and the synthetic control unit side-by-side, as is standard practice in Synthetic Control methodology, we would observe systematic discrepancies. This would then seem to indicate poor covariate balance, suggesting that the strong pre-treatment outcome fit was spurious. By design of this thought experiment, however, we know this is not the case—the synthetic control was constructed accounting for all covariates. Rather, this discrepancy is explained by the difference in absolute values of the outcome. We expect a higher outcome value to correspond to proportionally higher covariate values. Consequently, the standard procedure of examining the difference in absolute values of the covariates is not applicable when $\alpha \neq 0$. The workaround I propose is to display $X_1^{\Delta}$ and

$X_0^{\Delta} W$ in the covariate table, instead of $X_1$ and $X_0$. That is, we should evaluate the synthetic control based on how closely it is able to reconstruct the average pre-treatment change of the treated unit in the outcome and observable covariates. Consequently, the solution is to use a table just like Table 1 presented in section 2.4, but where the first two columns display average change of variables, as opposed to average values. The logic for including a third column showing the pairwise discrepancies proposed in section 2.4 is unaffected by any modifications of this novel estimator, as bias still scales with pairwise discrepancies. As such, the rationale for the inclusion of a third column displaying the weighted sum of pairwise differences still holds.[4]

---

[4] *Although it might be more appropriate to normalize the values with respect to the pre-treatment values of the treated unit, such that the entries in the third column can be interpreted as the average percentage difference between untreated units in the synthetic control and the treated unit. This is because the two first columns show the average change, as opposed to average value, and can thus not be used as indicators of the magnitude of the WMAPE.*

# 5. Empirical evaluation

This section aims to evaluate the penalized differenced synthetic control method against pre-existing methods, both the original synthetic control and the penalized synthetic control, using real data. Because of the fundamental problem of causal inference, we may never observe a true counterfactual against which to compare our models. To overcome this limitation, I fit the models to units where the estimated quantity is the observed outcome, namely untreated units. Specifically, I isolate one untreated unit in a dataset used in a paper applying the synthetic control method, pretend it was subjected to a treatment at some time $T_0$, and fit a synthetic control unit to it using the remaining untreated units as the donor pool. Because the untreated unit never received the treatment, the true treatment effect is zero. I repeat this procedure for each untreated unit. Consequently, we can compare models based on how close to ground truth estimates are.

## 5.1 Data

I use data from three seminal synthetic control studies: Abadie and Gardeazabal's (2003) study on the economic impact of terrorism in the Basque country; Abadie, Diamond and Hainmueller's (2010) study on the impact on California cigarette sales of Proposition 99, a tobacco control program; and Abadie, Diamond and Hainmueller's (2015) study on the impact of the German reunification on West Germany's economy. The Basque dataset contains 17 untreated units and 13 variables, including the outcome variable. The corresponding numbers for the California study are 38 and 5, and 16 and 8 for the West Germany study.[5]

## 5.2 Methodology

Four synthetic control estimators were compared in this empirical evaluation. The ordinary synthetic control, synthetic control with the penalization term from equation (7), differenced synthetic control i.e. penalized differenced synthetic control with $\lambda=0$, and penalized differenced synthetic control.

The following procedure was repeated for each synthetic control variant and dataset:

1. For each untreated unit, fit the synthetic control estimator using remaining untreated units as the control group and the same pre-treatment period as in the original study.

2. Record the root mean square prediction error (RMSPE) of the synthetic control estimate in the pre-treatment and the post-treatment period, respectively.

3. Record the sparsity of the resultant synthetic control, measured as the proportion of units in the control group that were assigned a non-zero weight in the synthetic control solution, **W**.

I choose the same pre-treatment period as used in the original study, even though none of the units in our empirical evaluation underwent the same treatment as the original treated unit. The reason is that each untreated unit was, as part of the original study, vetted to ensure the assumptions detailed in section 2.2 hold for that time period (Abadie and Gardeazabal, 2003; Abadie et al., 2010; Abadie et al., 2015).

---

[5] All three datasets are available at:
https://github.com/OscarEngelbrektson/SyntheticControlMethods/tree/master/examples/datasets

## 5.3 Results

| Dataset | Method | Pre RMSPE mean | Post RMSPE mean | Sparsity mean |
|---|---|---|---|---|
| basque_data | DSC | 0.18 | 0.42 | 0.19 |
| | DSC_pen | 0.17 | 0.36 | 0.20 |
| | SC | 0.37 | 0.61 | 0.20 |
| | SC_pen | 0.32 | 0.52 | 0.15 |
| german_reunification | DSC | 357.76 | 1642.52 | 0.22 |
| | DSC_pen | 374.51 | 1747.07 | 0.22 |
| | SC | 642.60 | 1934.76 | 0.14 |
| | SC_pen | 566.66 | 1841.99 | 0.21 |
| smoking_data | DSC | 6.66 | 15.21 | 0.28 |
| | DSC_pen | 6.02 | 12.90 | 0.11 |
| | SC | 8.83 | 13.76 | 0.06 |
| | SC_pen | 8.15 | 13.16 | 0.05 |

*Table 2.* Table showing the pre-treatment RMSPE, post-treatment RMSPE and sparsity of each model for each dataset. Smoking_data is the California study from Abadie et al. (2015). SC stands for synthetic control and DSC stands for differenced synthetic control. The suffix _pen indicates the penalized version of the estimator. The rank of each model, relative to the other models for that specific statistic and dataset, is color coded from dark blue (best) to dark red (worst). For example, the top-left cell is coloured light-blue because the differenced synthetic control method had the second lowest Pre RMSPE on the Basque dataset.[6]

---

[6] *In terms of sparsity, the original synthetic control and its penalized variant are consistently more sparse than the differenced estimators. That said, sparsity has no intrinsic value, only instrumental value in*

Analyzing Table 2, a few patterns emerge. Firstly, the pre-treatment RMSPE is consistently lower for both the  differenced synthetic control  estimators, as compared to the synthetic control estimators—almost a factor of two on the Basque dataset. This is not surprising as the estimator is more flexible by design, as described in section 3.2, thereby allowing a better fit on the period explicitly optimized on.

In terms of post-treatment RMSPE, which is arguably the most important metric as it is used to compute effect estimates, the penalized  differenced synthetic control method has the most attractive performance. It has the lowest statistic for the Basque dataset, by a large margin vis-a-vis both synthetic control estimators, and the Proposal 99 dataset. It comes in at a close second on the German reunification dataset after the differenced synthetic control method. The the differenced synthetic control method, however, ranked last on the smoking dataset detracting from its attractiveness. These results indicate that the penalization term has the intended effect of decreasing bias in the post-treatment period. Indeed, this is true also for the synthetic control method vis-a-vis the penalized synthetic control method for all datasets, both for pretreatment and posttreatment RMSPE.

Lastly, the penalized version of the synthetic control outperformed the original synthetic control on all metrics across all datasets, lending further credence to the claim that pairwise

---

facilitating qualitative interpretation and evaluation of estimates. As such, a higher proportion of units assigned a non-zero weight in the synthetic control is only a problem if it reaches a level where the interpretability of estimates is made difficult. Viewed from this lens, all estimators evaluated are sparse–although the unpenalized  differenced synthetic control is clearly worse, bordering on uninterpretability in the California study, as there are 38 untreated units. In general, we should expect the differenced synthetic control method to be less sparse, as it is more likely to have a non-unique solution for the optimal synthetic control (Abadie and L'Hour, 2020). Indeed, the rationale for the differencing is to provide a multiplicity of potential solutions for the penalization term to choose from.

differences should be considered and minimized for the sake of bias, even if uniqueness or sparsity is not a concern.

Overall, the penalized differenced synthetic control compares favourably to the other methods in terms of the quality of fit both in pre- and post-treatment periods, without sacrificing the interpretability of estimates.

## 6. Conclusion

In this paper, I show that synthetic control estimators are generally biased if the functional relationship between outcome and covariates is non-linear. I show that this bias increases with the weighted sum of absolute pairwise differences between the untreated units in the synthetic control and the treated unit. As such, I argue that it should be actively minimized through a penalization term, and considered in evaluating the pre-treatment fit of synthetic controls, even in the cases where uniqueness is not an issue. Further, I introduce a novel synthetic control estimator which mitigates this bias. It has increased flexibility, which increases the number of potential synthetic untreated units that provide strong pre-treatment fit, and systematically selects the synthetic control which will minimize bias. I present a data-driven estimation process, including the data-driven choice of penalization parameter, whilst maintaining compatibility with inferential methods introduced in Abadie et al. (2010). I show that the novel estimator performs well on real data and that penalization improves performance of synthetic control estimators.

I see three primary avenues for future research. Firstly, investigating the performance consequences of adding penalization to other synthetic control estimators proposed in the literature, such as the synthetic difference-in-differences estimator from Arkhangelsky et al. (2019). Secondly, extending the theoretical analysis presented in Section 3.3 to derive an upper bound on the bias of Synthetic Controls under non-linear generative models. Thirdly, further analysis on the properties of relaxed and penalised synthetic controls.

# References

Abadie, A., Diamond, A., Hainmueller, J. (2010). Synthetic Control Methods for Comparative Case Studies: Estimating the Effect of California's Tobacco Control Program. Journal of the American Statistical Association.

Retrieved from

https://economics.mit.edu/files/11859

Abadie, A., Diamond, A., Hainmueller, J. (2015). Comparative Politics and the Synthetic Control Method. American Journal of Political Science.

Retrieved from

https://web.stanford.edu/~jhain/Paper/AJPS2015a.pdf

Abadie, A., Gardeazabal, J. (2003). The Economic Costs of Conflict: A Case Study of the Basque Country. American Economic Review.

Retrieved from

https://economics.mit.edu/files/11870

Abadie, A., L'Hour, J. (2020). A Penalized Synthetic Control Estimator for Disaggregated Data. Working Paper.

Retrieved from

http://economics.mit.edu/files/18642

Arkhangelsky, D., Athey, S., Hirshberg, A. D., Imbens, W. G., Wager, S. (2019). Synthetic Difference in Differences. NBER working paper series.

Retrieved from

https://www.nber.org/papers/w25532.pdf

Athey, S., Imbens, G. (2017). The State of Applied Econometrics: Causality and Policy Evaluation. Journal of Economic Perspectives.

Retrieved from

https://faculty.smu.edu/millimet/classes/eco7377/papers/athey%20imbens%202017.pdf

Ben-Michael, E,. Feller, A., Rothstein, J. (2019). The Augmented Synthetic Control Method. ArXiv preprint.

Retrieved from

https://arxiv.org/pdf/1811.04170.pdf

Chernuzhukov, V., Wüthrich, K., Zhu, Y. (2020). Practical and robust t-test based inference for synthetic control and related methods.

Retrieved from

https://arxiv.org/pdf/1812.10820.pdf

Cotton, P. (2021). Popular Python Time Series Packages. Microprediction.

Retrieved from

https://www.microprediction.com/blog/popular-timeseries-packages

Cunningham, S. (2018). Causal Inference: Mixtape. tufte-latex.googlecode.com

Retrieved from

http://scunning.com/cunningham_mixtape.pdf

Cunningham, S., Kang, S. (2018). Incapacitating Drug Users: Evidence from Prison Construction. Baylor University.

Retrieved from

http://scunning.com/prison_booms_and_drugs_20.pdf

Engelbrektson, O. (2021). SyntheticControlMethods: A Python package for causal inference using Synthetic Controls. Github.

    Retrieved from

    https://github.com/OscarEngelbrektson/SyntheticControlMethods

Ferman, B., Pinto, C. (2020). Synthetic controls with imperfect pre-treatment fit. arXiv preprint.

    Retrieved from

    https://arxiv.org/pdf/1911.08521.pdf

Jones, B., Barrows, S. (2019). Uber's Synthetic Control. PyData Conference Amsterdam 2019.

    Retrieved from

    https://www.youtube.com/watch?v=j5DoJV5S2Ao&ab_channel=PyData

Shimizu, S., Hoyer, O. P., Hyvärinen, A., Kerminen, A. (2006). A Linear Non-Gaussian Acyclic Model for Causal Discovery. Journal of Machine Learning Science.

    Retrieved from

    https://www.cs.helsinki.fi/group/neuroinf/lingam/JMLR06.pdf

## Appendix A: Complete algebra for Section 2.3

This Appendix is included for completeness and shifting some non-crucial algebra away from the main text. Given the factor model described by equation (4), $\mathbf{X_1} = \mathbf{X_0}\mathbf{W}$ implies that that

$Y_{1t}^{N} = \sum\limits_{i=2}^{J+1} w_i Y_{it}$ and $Z_1 = \sum\limits_{i=2}^{J+1} w_i Z_{it}$ for all t $\leq$ T$_0$. Substituting $Y_{it}^{N} = \theta_t Z_i^2$ from equation

(4) and dividing by $\theta_t$, we get there must be a solution $\mathbf{W}$ such

$$Z_1 = \sum\limits_{i=2}^{J+1} w_i Z_i \quad \text{and} \quad Z_1^2 = \sum\limits_{i=2}^{J+1} w_i Z_i^2$$

Squaring the first equation, we get identical left hand sides of both equations. Consequently, this is implies

$$\left(\sum\limits_{i=2}^{J+1} w_i Z_i\right)^2 = \sum\limits_{i=2}^{J+1} w_i Z_i^2$$

Which is generally impossible given the convexity constraints on $\mathbf{W}$. Indeed, for any non-linear transformation $\Phi(\bullet)$ it is generally true that,

$$\Phi\left(\sum\limits_{i=2}^{J+1} w_i Z_i\right) \neq \sum\limits_{i=2}^{J+1} w_i \Phi(Z_i)$$

As non-linear operators do not associate.